# Quantum Holography

**Ayman F. Abouraddy, Bahaa E. A. Saleh, Alexander V. Sergienko and Malvin C. Teich**
*Boston University, Quantum Imaging Laboratory, Departments of Electrical & Computer Engineering and Physics, 8 Saint Mary's Street, Boston, Massachusetts 02215, USA*
*besaleh@bu.edu*

*http://www.bu.edu/qil*

**Abstract:** We propose to make use of quantum entanglement for extracting holographic information about a remote 3-D object in a confined space which light enters, but from which it cannot escape. Light scattered from the object is detected in this confined space entirely without the benefit of spatial resolution. Quantum holography offers this possibility by virtue of the fourth-order quantum coherence inherent in entangled beams.

## 1. Introduction

We consider the use of quantum entanglement [1], which gives rise to 'spooky actions at a distance' in Einstein's words [2], for extracting holographic information [3,4] about a remote 3-D object concealed in an integrating sphere. Quantum holography makes use of entangled-photon pairs [5,6], one of which one scatters from the remote object while the other is locally manipulated using conventional optics that offers full spatial resolution. Remarkably, the underlying entanglement permits the measurement to yield coherent holographic information about the remote object. Quantum holography offers this possibility by virtue of the fourth-order quantum coherence inherent in entangled beams; indeed, it can be implemented despite the fact that conventional second-order coherence, required for ordinary holography, is absent. Classical holography cannot achieve this. Belinskii and Klyshko [8] constructed a two-photon analog of classical holography, although they provided no analysis. The configuration presented here makes use of entanglement to transcend the capabilities of classical holography.

Specifically, consider a 3-D object placed within a chamber that has an opening through which light enters but does not escape, as illustrated in Fig. 1. Coated with a photosensitive surface, the wall of the chamber serves as an integrating sphere that converts *any* photon reaching it into a photoevent. The chamber therefore serves as a photon bucket that indiscriminately detects the arrival of photons at any point on its surface, whether scattered or not, but is totally incapable of discerning the location at which the photon arrives.

Classically it is impossible to construct a hologram of the 3-D object in this configuration, whatever the nature of the light source or the construction of the imaging system. This is because optical systems that make use of classical light sources, even those that involve scanning and time-resolved imaging, are incapable of resolving the ambiguity of positions from which the photons are scattered; they therefore cannot be used to form a coherent image suitable for holographic reconstruction.

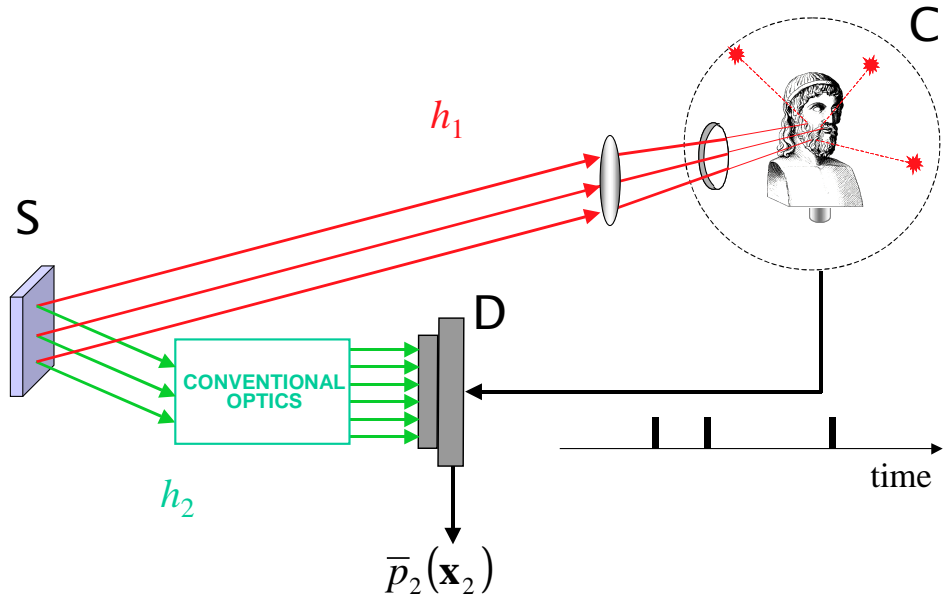

**Figure 1:** Quantum holography. S is a source of entangled-photon pairs. C is a (remote) single-photon-sensitive integrating sphere that comprises the wall of the chamber concealing the hidden object (bust of Plato). D is a (local) 2-D single-photon-sensitive scanning or array detector. $h_1$ and $h_2$ represent the optical systems that deliver the entangled photons from S to C and D, respectively. The quantity $\overline{p}_2(\mathbf{x}_2)$ is the marginal coincidence rate, which is the hologram of the concealed object. Thin and thick lines represent optical and electrical signals, respectively.

## 2. Method

The implementation of quantum holography makes use of entangled-photon beams generated, for example, by the process of spontaneous optical parametric down-conversion [6-12] from a second-order nonlinear crystal illuminated by a pump laser. As shown in Fig. 1, one beam from the source S enters the chamber opening and is scattered from the object, yielding a single sequence of photoevents from the integrating sphere C. The other beam is transmitted through a conventional optical system and detected using a single-photon-sensitive scanning (or array) detector D. The information registered by the two detectors, in the form of coincidence counts, is sufficient to extract coherent information about the 3-D object that is suitable for holographic reconstruction.

Let S be a planar two-photon source emitting photons in a pure entangled quantum state [8]

$$|\Psi\rangle = \iint_S d\mathbf{x} d\mathbf{x}' \varphi(\mathbf{x}) \delta(\mathbf{x} - \mathbf{x}') |1_\mathbf{x}\rangle \otimes |1_{\mathbf{x}'}\rangle, \quad (1)$$

where $\mathbf{x} \in S$, $|1_\mathbf{x}\rangle = \frac{1}{(2\pi)^2} \int d\mathbf{k} \exp(i\mathbf{k} \cdot \mathbf{x}) |1_\mathbf{k}\rangle$ is a position representation of the single-photon state in terms of the familiar Fock state $|1_\mathbf{k}\rangle$ of the mode with wave vector $\mathbf{k}$, $\delta$ is the Dirac delta function, and the state probability amplitude $\varphi(\mathbf{x})$ is normalized such that $\int_S d\mathbf{x} |\varphi(\mathbf{x})|^2 = 1$. Here $|\varphi(\mathbf{x})|^2$ represents the probability density that a photon pair is emitted from point $\mathbf{x}$ in the source plane. As a consequence of the state in Eq. (1), each photon is individually in a mixed state (described by the density operator $\hat{\rho} = \int_S d\mathbf{x} |\varphi(\mathbf{x})|^2 |1_\mathbf{x}\rangle\langle 1_\mathbf{x}|$) that exhibits no second-order coherence [9], as is required in traditional holography. This entangled state may be generated, for example, by spontaneous parametric down-conversion from a thin crystal, in which case $\varphi(\mathbf{x})$ represents the spatial distribution of the pump field [8].

Of the two photons generated by the source S, the one directed through the opening of the chamber may (or may not) be scattered from the object and impinges on the chamber wall at position $\mathbf{x}_1 \in C$, where C represents the set of points on the chamber wall. The optical system between the source and the chamber, idealized as a simple lens in Fig. 1, as well as everything inside the chamber including the object, is assumed to be linear and is characterized by an impulse response function $h_1(\mathbf{x}_1, \mathbf{x})$. The other photon is transmitted through a linear optical system characterized by an impulse response function $h_2(\mathbf{x}_2, \mathbf{x})$, where $\mathbf{x}_2 \in D$, the single-photon-sensitive scanning (or array) detector.

The photon coincidence rate at points $\mathbf{x}_1$ and $\mathbf{x}_2$ is described by a probability density $p(\mathbf{x}_1, \mathbf{x}_2)$ given by[7]

$$p(\mathbf{x}_1, \mathbf{x}_2) \propto \left| \int_S d\mathbf{x} \varphi(\mathbf{x}) h_1(\mathbf{x}_1, \mathbf{x}) h_2(\mathbf{x}_2, \mathbf{x}) \right|^2. \quad (2)$$

The form of Eq. (2) suggests that the function $p(\mathbf{x}_1, \mathbf{x}_2)$ may be regarded as the coherent image of a point $\mathbf{x}_1 \in C$ formed through an optical system represented by the following cascade (see Fig. 1): propagation through $h_1$ in the reverse direction toward the source (from $\mathbf{x}_1$ to $\mathbf{x}$), modulation by $\varphi(\mathbf{x})$ at the source, and subsequent transmission from the source through the system $h_2$ to the point $\mathbf{x}_2$. Equation (2) may therefore be written symbolically as follows: $p(\mathbf{x}_1,\mathbf{x}_2) \propto |h_2 * \varphi \cdot h_1|^2$, where $*$ represents transmission through a linear system (convolution in the shift-invariant case) and $\cdot$ represents multiplication or modulation. The expression is to be read in reverse order, from right to left, as is the custom in operator algebra.

Since we have no knowledge of the detection points $\mathbf{x}_1 \in \mathsf{C}$ on the chamber wall ($\mathsf{C}$ is a bucket detector) we must integrate over $\mathsf{C}$, whereupon the coincidence rate in Eq. (2) becomes

$$\bar{p}_2(\mathbf{x}_2) = \int_\mathsf{C} d\mathbf{x}_1 p(\mathbf{x}_1,\mathbf{x}_2) \propto \iint_S d\mathbf{x} d\mathbf{x}' \varphi(\mathbf{x}) \varphi^*(\mathbf{x}') h_2(\mathbf{x}_2,\mathbf{x}) h_2^*(\mathbf{x}_2,\mathbf{x}') g_1(\mathbf{x},\mathbf{x}'), \qquad (3)$$

with $g_1(\mathbf{x},\mathbf{x}') = \int_\mathsf{C} d\mathbf{x}_1 h_1(\mathbf{x}_1,\mathbf{x}) h_1^*(\mathbf{x}_1,\mathbf{x}')$. The function $\bar{p}_2(\mathbf{x}_2)$ is therefore the marginal probability density of detecting one photon at $\mathbf{x}_2$ and another at any point $\mathbf{x}_1 \in \mathsf{C}$. In spite of this integration, it is clear from Eq. (3) that $\bar{p}_2(\mathbf{x}_2)$ contains information about the system $h_1$, and therefore about the object, via the function $g_1$. The function $\bar{p}_2(\mathbf{x}_2)$ is the incoherent superposition of many coherent images of the form given in Eq. (2), originating from all points of $\mathsf{C}$. This is therefore a modal expansion of a partially coherent system [12].

## 3. Example: Scattering objects

To illustrate the principle, let us consider two samples, in turn: a single point scatterer and a collection of such scatterers. These results are readily generalized to an arbitrary object. Consider a single static scatterer located at the point $\mathbf{x}^{(1)}$ inside $\mathsf{C}$ as depicted in Fig. 2. The system $h_1$ comprises two contributions. The first is a direct path to the chamber wall, represented by the system $h^{(0)}$. The second is a scattering path to the chamber wall, represented by the illumination system $h_I^{(1)}$ that directs light to the point scatterer, the fraction of the field that is scattered (the complex scattering strength) $\varepsilon(\mathbf{x}^{(1)})$, and the system $h^{(1)}$ that carries light from the scatterer to the chamber wall. These two paths are mutually coherent, so that the probability amplitudes of the two paths are added, thereby leading to

$$h_1(\mathbf{x}_1,\mathbf{x}) = h^{(0)}(\mathbf{x}_1,\mathbf{x}) + h^{(1)}(\mathbf{x}_1,\mathbf{x}^{(1)}) \varepsilon(\mathbf{x}^{(1)}) h_I^{(1)}(\mathbf{x}^{(1)},\mathbf{x}). \qquad (4)$$

Substituting Eq. (4) into Eq. (3) yields the marginal coincidence rate

$$\bar{p}_2(\mathbf{x}_2) \propto \bar{p}_2^{(0)}(\mathbf{x}_2) + \bar{p}_2^{(1)}(\mathbf{x}_2) + 2\,\mathrm{Re}\{\varepsilon(\mathbf{x}^{(1)}) r(\mathbf{x}_2,\mathbf{x}^{(1)}) q(\mathbf{x}_2,\mathbf{x}^{(1)})\}, \qquad (5)$$

with

$$\bar{p}_2^{(1)}(\mathbf{x}_2) = |\varepsilon(\mathbf{x}^{(1)})|^2 \beta(\mathbf{x}^{(1)},\mathbf{x}^{(1)}) |q(\mathbf{x}_2,\mathbf{x}^{(1)})|^2, \qquad (6)$$

$$\beta(\mathbf{x}^{(1)},\mathbf{x}^{(1)}) = \int d\mathbf{x}_1 |h^{(1)}(\mathbf{x}_1,\mathbf{x}^{(1)})|^2, \qquad (7)$$

$$q(\mathbf{x}_2,\mathbf{x}^{(1)}) = \int d\mathbf{x} \varphi(\mathbf{x}) h_2(\mathbf{x}_2,\mathbf{x}) h_I^{(1)}(\mathbf{x}^{(1)},\mathbf{x}), \qquad (8)$$

$$r(\mathbf{x}_2, \mathbf{x}^{(1)}) = \int d\mathbf{x}\, \varphi^*(\mathbf{x}) f(\mathbf{x}^{(1)}, \mathbf{x}) h_2^*(\mathbf{x}_2, \mathbf{x}), \tag{9}$$

$$f(\mathbf{x}^{(1)}, \mathbf{x}) = \int d\mathbf{x}_1\, h^{(0)*}(\mathbf{x}_1, \mathbf{x}) h^{(1)}(\mathbf{x}_1, \mathbf{x}^{(1)}). \tag{10}$$

Equation (5) is the sum of three terms, which may be elucidated by referring to Fig. 2 that depicts the Feynman-like paths of the various probability amplitudes: (1) The first term $\bar{p}_2^{(0)}(\mathbf{x}_2)$ is the marginal coincidence rate in absence of the scatterer. It is identical to that in Eq. (3) with $h_1$ replaced by $h^{(0)}$. This term represents the direct path in Fig. 2. (2) The second term $\bar{p}_2^{(1)}(\mathbf{x}_2)$ is the marginal coincidence rate arising from the scatterer alone, and is represented by the scattering path in Fig. 2. (3) The third term represents interference between these two paths, and is therefore the term of interest for quantum holography. It is the fourth-order analog of second-order interference in Gabor's original conception of holography [3,4]. One may represent the functions $r(\mathbf{x}_2, \mathbf{x}^{(1)})$ and $q(\mathbf{x}_2, \mathbf{x}^{(1)})$, which are defined in Eqs. (8) and (9), respectively, by the symbolic relations: $r = h_2^* * \varphi^* \cdot h^{(0)*} * h^{(1)}$ and $q = h_2 * \varphi \cdot h_I^{(1)}$. In other words, $r(\mathbf{x}_2, \mathbf{x}^{(1)})$ is the image formed by a point at the location of the scatterer $\mathbf{x}^{(1)}$ through a cascade of the systems $h^{(1)}$ (traveling forward) and $h^{(0)}$ (traveling backward), followed by modulation by $\varphi$, and finally traveling forward through the system $h_2$ to the point $\mathbf{x}_2$. This is the term that includes the holographic information. The quantity $q(\mathbf{x}_2, \mathbf{x}^{(1)})$, by which $r$ is multiplied in Eq. (5), is the image of a point at $\mathbf{x}^{(1)}$ traveling backward through $h_I^{(1)}$, followed by modulation by $\varphi$ and then forward propagation through $h_2$. If the optical system is designed such that $h_I^{(1)}$ is uniform over the area of interest, then $q$ is independent of $\mathbf{x}^{(1)}$ and becomes unimportant. Note that integration over the area of the chamber is essential for achieving quantum holography. Thus a point detector, for example [8], cannot be used for this purpose by virtue of Eqs. (8) - (10). Furthermore, ray tracing techniques, such as those used in used in Ref. [13] in connection with geometric optics of entangled-photon beams, cannot be used for characterizing this interference effect.

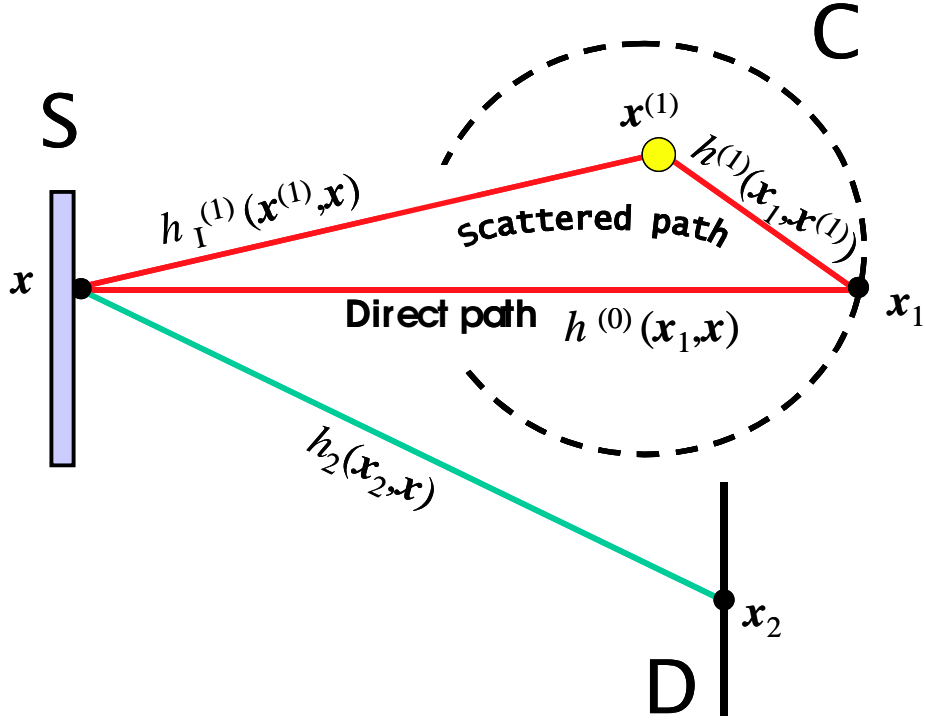

Figure 2. Quantum holography of a single point scatterer located at point $\mathbf{x}^{(1)}$ inside C. All quantities are defined in the text.

Consider now the case when $N$ static scatterers are located at positions $\mathbf{x}^{(j)}, j = 1..N$, inside C, whereupon the impulse response function $h_1$ becomes

$$h_1(\mathbf{x}_1, \mathbf{x}) = h^{(0)}(\mathbf{x}_1, \mathbf{x}) + \sum_{j=1}^{N} h^{(j)}(\mathbf{x}_1, \mathbf{x}^{(j)}) \varepsilon(\mathbf{x}^{(j)}) h_1^{(j)}(\mathbf{x}^{(j)}, \mathbf{x}), \qquad (11)$$

which is a generalization of Eq. (4). The marginal coincidence rate in this case, obtained by substituting Eq. (11) into Eq. (3), becomes

$$\bar{p}_2(\mathbf{x}_2) \propto \bar{p}_2^{(0)}(\mathbf{x}_2) + \bar{p}_2^{(\Sigma)}(\mathbf{x}_2) + 2\,\mathrm{Re}\left\{ \sum_{j=1}^{N} \varepsilon(\mathbf{x}^{(j)}) r(\mathbf{x}_2, \mathbf{x}^{(j)}) q(\mathbf{x}_2, \mathbf{x}^{(j)}) \right\}, \qquad (12)$$

which is a generalization of Eq. (5). Here

$$\bar{p}_2^{(\Sigma)}(\mathbf{x}_2) = \sum_{j=1}^{N} \bar{p}_2^{(j)}(\mathbf{x}_2) + 2\,\mathrm{Re}\left\{ \sum_{j=1, i=j+1}^{N} \varepsilon(\mathbf{x}^{(j)}) \varepsilon^*(\mathbf{x}^{(i)}) \beta(\mathbf{x}^{(j)}, \mathbf{x}^{(i)}) q(\mathbf{x}_2, \mathbf{x}^{(j)}) q^*(\mathbf{x}_2, \mathbf{x}^{(i)}) \right\}, \qquad (13)$$

$$\bar{p}_2^{(j)}(\mathbf{x}_2) = \left| \varepsilon(\mathbf{x}^{(j)}) \right|^2 \beta(\mathbf{x}^{(j)}, \mathbf{x}^{(j)}) \left| q(\mathbf{x}_2, \mathbf{x}^{(j)}) \right|^2, \qquad (14)$$

$$\beta(\mathbf{x}^{(j)}, \mathbf{x}^{(i)}) = \int d\mathbf{x}_1 h^{(j)}(\mathbf{x}_1, \mathbf{x}^{(j)}) h^{(i)*}(\mathbf{x}_1, \mathbf{x}^{(i)}), \qquad (15)$$

with all other quantities as previously defined. Once again the marginal coincidence rate, given in Eq. (12), is the sum of three terms analogous to those in Eq. (5). The second term $\overline{p}_2^{(\Sigma)}(\mathbf{x}_2)$, that due to the scatterers alone, includes the sum of the contributions of each scatterer independently, and terms resulting from interference amongst the scatterers. The third term in Eq. (12) includes the holographic information. The results can then be generalized to any object by replacing the discrete summation in Eq. (11) by an integral. The results also apply to objects that do not scatter.

The image obtained from the marginal coincidence rate $\overline{p}_2(\mathbf{x}_2)$ is holographic by virtue of Eq. (12). This equation has the structure of a conventional hologram obtained by illuminating the scatterers with coherent light through a composite system involving the optics of both beams, with the state probability amplitude $\varphi(\mathbf{x})$ serving as an effective coherent aperture. This result follows from the duality between entanglement and coherence [8].

*Thus $\overline{p}_2(\mathbf{x}_2)$ is a hologram of the 3-D object concealed in the chamber. It may then be recorded on a 2-D photographic plate and viewed with ordinary light in the usual fashion of holographic reconstruction.*

### 4. Conclusion

The remarkable possibility of quantum holography is attained by virtue of a light beam that itself does not illuminate the object, but is entangled with the beam that does, and is detected with full spatial resolution. Although each of the beams is, by itself, incoherent, and therefore not capable of conventional interference, and although the integrating sphere provides no spatial resolution whatsoever, the quantum entanglement permits interference and hence offers the possibility of holography. This surprising and purely quantum result cannot be achieved by using optical beams generated by a classical source, even if they possess the strongest possible classical correlation [9].

### 7. Acknowledgements

This work was supported by the US National Science Foundation; by the Center for Subsurface Sensing and Imaging Systems (CenSSIS), an NSF engineering research center; and by the David & Lucile Packard Foundation.